\documentclass[sigconf,natbib=false]{acmart}

\AtBeginDocument{%
  }

\copyrightyear{2026}
\acmYear{2026}
\setcopyright{cc}
\setcctype{by}
\acmConference[SAC '26]{The 41st ACM/SIGAPP Symposium on Applied Computing}{March 23--27, 2026}{Thessaloniki, Greece}
\acmBooktitle{The 41st ACM/SIGAPP Symposium on Applied Computing (SAC '26), March 23--27, 2026, Thessaloniki, Greece}
\acmDOI{10.1145/3748522.3779882}
\acmISBN{979-8-4007-2294-3/2026/03}

\RequirePackage[
  datamodel=acmdatamodel,
  style=acmnumeric,
  ]{biblatex}

\usepackage{xspace}
\usepackage{listings}

\newcommand\code[1]{{\tt \small #1}}

\newcommand{\donothing}{Do-Nothing\xspace}
\newcommand{\manualpollute}{Manual-Pollute\xspace}

\lstnewenvironment{inlinelisting}{
\lstset{language=Java,
numbers=none,
basicstyle=\fontsize{8}{8}\selectfont\ttfamily
}}{}

\colorlet{punct}{red!60!black}
\definecolor{background}{HTML}{EEEEEE}
\definecolor{delim}{RGB}{20,105,176}
\colorlet{numb}{magenta!60!black}

\begin{document}

\title{Misleading Microbenchmarks on the Java Virtual Machines}

\author{Filippo Schiavio}
\email{filippo.schiavio@usi.ch}
\orcid{}
\affiliation{%
  \institution{Univertistà della Svizzera italiana}
  \city{Lugano}
  \country{Switzerland}
}

\author{Lubomír Bulej}
\email{bulej@d3s.mff.cuni.cz}
\orcid{}
\affiliation{%
  \institution{Charles University}
  \city{Praha}
  \country{Czech Republic}
}

\author{Walter Binder}
\email{walter.binder@usi.ch}
\orcid{}
\affiliation{%
  \institution{Univertistà della Svizzera italiana}
  \city{Lugano}
  \country{Switzerland}
}

\begin{abstract}
Developers often use microbenchmarks to choose the most performant implementation of a method or a class. On the Java Virtual Machine (JVM), this is commonly done using the Java Microbenchmark Harness (JMH) which addresses common pitfalls of measuring code performance on the JVM. However, even using JMH guidelines cannot overcome the fundamental issue of context.

Microbenchmarks inherently execute code in isolation, without interference from other application code competing for CPU resources, such as cache or branch-predictor capacity. On managed runtimes with tiered dynamic compilation, such as the JVM, the speculative, profile-driven nature of compilation decisions means that code performance is highly dependent on profiles collected during early execution. Because profiles usually include also branch probabilities and receiver types (besides code hotness metrics), a badly designed microbenchmark may cause the JVM to collect an unrealistic profile, resulting in aggressive, yet misleading, optimizations, that would not occur in a real application.

In this paper, we demonstrate how using microbenchmarks under conditions that induce the JVM to collect unrealistic profiles yields misleading results despite following existing guidelines. We also extend these guidelines by suggesting actions to make the microbenchmark results more representative.
\end{abstract}

\ccsdesc[500]{Software and its engineering~Software performance}
\ccsdesc[300]{Software and its engineering~Runtime environments}
\ccsdesc[300]{Software and its engineering~Just-in-time compilers}

\keywords{Benchmarking, Microbenchmarks, JVM, Misleading Performance, Profile-driven Dynamic Compiler Optimizations}

\maketitle

\section{Introduction}\label{sec:intro}
The process of software-performance optimization is commonly driven by benchmarks, i.e., a certain implementation is evaluated against an existing baseline by measuring different metrics, such as execution time and memory consumption. 
This process is successful when the alternative implementation results in an optimization, i.e., a code version that shows better metrics than its baseline on certain benchmarks, or on a relevant subset of use cases. 

Benchmarks are usually large codebases inspired by real-world applications. Widely used benchmark suites such as SPEC~\cite{specjvm98,specjvm2008}, DaCapo~\cite{dacapo,dacapo-chopin} and Renaissance~\cite{renaissance} adopt this methodology. 
However, developers commonly need to optimize a particular implementation, e.g., a single method or a data structure in a certain library. 
If the benchmarks do not spend a significant amount of time on the evaluated functionality, performance differences will likely go unnoticed, making it very difficult to assess performance of the optimized functionality. 
To solve this problem, developers commonly make use of so-called microbenchmarks, which is the canonical approach used to drive the optimization process of performance-critical small code fragments, evaluating them in isolation.
In the context of the Java Virtual Machine (JVM), the established example of such a framework is Java Microbenchmark Harness (JMH)~\cite{jmh}, which is widely used, even by the language implementers. 

Microbenchmarks typically rely on input data that is synthetic, either randomly generated or cherry-picked by the developers. 
In particular, JMH allows developers to provide an easy-to-use annotation-based approach for expressing the input parameters for each benchmark.  
Remarkably, with the goal of minimizing perturbation of measurements and therefore the variance among multiple runs, JMH runs each benchmark with a configuration of parameters in isolation.
To this end, for each configuration of parameters, JMH spawn a new JVM process and collects the performance metrics of the given parameter list.
In this paper, we will say that such an execution strategy evaluates the implementation of interest in a \emph{sterile environment}, i.e., the implementation is used only by a single method, i.e., the benchmarked one, and usually called by a single call site, passing always the same parameters, therefore exercising always the same code paths within the implementation. 
A sterile environment is clearly unlikely, if not impossible, to show up in a real-world application, since tipically a library is used in multiple code locations, and invoked with different parameters. 

In this paper we focus on the JVM as virtual machine and on execution time as performance metric.
We show multiple case studies where the sterile environment of microbenchmarks results in misleading performance behavior.
In particular, we show implementations that are often---or always---faster than a certain baseline when executed in a sterile environment but slower when executed in a realistic setting. 
To further support our claim, we show that even the performance of the very same implementation is substantially different if the implementation is used in a sterile environment or in a realistic setting.
Therefore, we conclude that a widely used approach in performance engineering may lead developers to derive wrong conclusions and therefore to make wrong choices during the software optimization process. 
Besides showing the unexpected performance behavior arising from microbenchmarking in a sterile environment, we analyze the root causes of this phenomenon and we provide concrete suggestions to developers and researchers to avoid it in their own benchmarks.

Our work makes the following contributions:
\begin{enumerate}
    \item We describe a common issue of microbenchmarks, where code is executed in a sterile environment, resulting in narrow profiles and unrealistic compiler optimizations.
    \item We show three case studies of misleading performance behaviors which could lead developers to wrong conclusions, and we analyze their root causes.
    \item We derive a list of hints for developers, researchers and practitioners to reduce the likelihood of misleading microbenchmarks.
\end{enumerate}

This paper is structured as follows:
Sec.~\ref{sec:background} introduces the background in the context of this work.
Sec.~\ref{sec:implication} describes the problem tackled in this work.
Sec.~\ref{sec:overview} presents an overview of our experimentation plan and describes the evaluation settings.
We describe our three case studies in Sections~\ref{sec:cs1}, \ref{sec:cs2}, and~\ref{sec:cs3}.
Finally, we discuss related work in Sec.~\ref{sec:relwork}, and Sec.~\ref{sec:end} concludes the paper.

\section{Background}\label{sec:background}
Here, we provide the necessary background covering the JVM execution model and tiered compilation (§~\ref{sec:background:execmodel}), commonly used profile-guided optimizations (§~\ref{sec:background:pgo}), important aspects of the JVM initialization (§~\ref{sec:background:init}), and the execution model of JMH benchmarks (§~\ref{sec:background:jmh}).

\subsection{The JVM Execution Model}\label{sec:background:execmodel}
Applications implemented in Java are compiled into a platform independent bytecode, which is then executed by a JVM.
Most of the JVM implementations execute Java bytecode by combining interpretation with just-in-time (JIT) compilation, using a so-called tiered execution strategy. 
As an example, with HotSpot, the most widely used JVM implementation, methods are initially interpreted, then selectively compiled by one of two JIT compilers---the client compiler (\texttt{C1}), designed to obtain fast compilation applying only few optimizations, and the server compiler (\texttt{C2}), designed to obtain very optimized code, at the cost of a slower compilation~. 
Indeed, since compilation takes place at runtime, only the methods that are detected as ``hot'' by the virtual machine (i.e., they are invoked many times or they contain loops that perform many iterations) are subject to compilation, particularly with \texttt{C2}.
Other JVMs, such as GraalVM~\cite{duboscq2013graal,wurthinger2013one}, provide alternative JIT compilers with more advanced optimizations, particularly GraalVM replaces the \texttt{C2} compiler with the \texttt{Graal} compiler.

\subsection{Profile-Guided Optimizations}\label{sec:background:pgo}
JIT compilers rely on runtime profiling data to guide compilation decisions, an approach called profile guided optimizations (PGO).  
Counters such as method invocation frequencies, branch probabilities, and type distributions are recorded during interpretation or tier-1 compilation (e.g., in \texttt{C1}), and later used to select methods to be compiled and also to drive optimization decisions.
For example, methods invoked frequently are candidates for inlining and specialization, while less frequently used code may remain interpreted.  
Therefore, the role of runtime profiling information is crucial for optimizing Java application and other languages executed on virtual machines with a JIT compiler (e.g., JavaScript on V8). 

Runtime profiling information are also used to implement \emph{speculative optimizations}.  
These optimizations assume that future executions will be consistent with the collected profiles, and they generate optimized code under that assumption.  
With these optimizations only part of the actual application is compiled (e.g., only a single branch or specialization for a single type), leading to much smaller code size and, in turn, to additional optimization possibilities, since optimizations performed by JITs are often limited by thresholds (commonly called budgets).
A prominent mechanism is the use of polymorphic inline caches~\cite{holzle1991optimizing}, which record the receiver types observed at virtual call sites and allow the JIT to inline or devirtualize method calls based on this information.  
If later executions violate the speculation (e.g., a new receiver type appears), the JVM performs \emph{deoptimization}, invalidating the compiled code and falling back to interpretation until a new compilation is triggered~\cite{holzle1992debugging}.  
Speculative optimizations are central to JVM performance but also mean that the quality of generated code is sensitive to the diversity of inputs and contexts under which methods are exercised.    

\subsection{JVM Initialization Effects}\label{sec:background:init}
Another crucial factor often overlooked in benchmarking is the JVM initialization phase.  
During startup, the JVM executes significant portions of the Java Class Library (JCL), e.g., during the static initialization~\cite{static-initialization-oracle} of classes used by the JVM initialization, i.e., before executing user code.  
Core classes such as \texttt{java.lang.String} and many collections in the \texttt{java.util} package are exercised long before user code or benchmarks begin.  
This initialization work not only executes Java code but also contributes profiling information and can even trigger JIT compilation early in the process.

\subsection{The Execution Model of JMH Benchmarks}\label{sec:background:jmh}
The de facto standard framework for writing microbenchmarks on the JVM is the Java Microbenchmark Harness (JMH)~\cite{jmh}.
JMH is developed and maintained as part of OpenJDK, and is widely adopted both in academia and industry, including by JVM language implementers themselves.\footnote{As an example, on the latest JDK-24 branch on GitHub at the time of writing (JDK-24+36), a search for all Java files importing the JMH \texttt{Benchmark} class results in 538 matches, (\href{https://github.com/search?q=repo\%3Aopenjdk\%2Fjdk+jmh.annotations.benchmark+path\%3A*.java&type=code&ref=jdk-24+36}{GitHub search}).}
JMH has become the reference tool for performance evaluation of Java methods thanks to its careful design, which addresses many of the pitfalls that arise when measuring performance on virtual machines.
JMH provides several mechanisms to reduce measurement noise, to improve the reproducibility of results and to simplify the benchmark implementation:
\begin{itemize}
    \item \textbf{Warmup iterations.} Benchmarks are first executed for a configurable number of warmup iterations, ensuring that measurements are collected only once the JVM completes its ergonomics process~\cite{ergonomics}, that is, the adaptive heuristics to tune the currently running applications have been applied (e.g., by the garbage collector and the JIT compiler), so that the application is executing at steady state.
    \item \textbf{Multiple forks.} Each benchmark configuration can be executed in multiple JVM forks (i.e., different processes started sequentially from scratch), thereby mitigating sources of non-determinism such as lucky or unlucky memory layouts, JIT compilation timing, and garbage collection scheduling.
    \item \textbf{Statistical rigor.} JMH automatically computes descriptive statistics (mean, median, standard deviation, confidence intervals, percentiles) across runs, helping developers to reason about variance and stability of results.
    \item \textbf{Isolation.} Each benchmark is executed in a fresh JVM process,  avoiding cross-benchmark interference, such as classloader pollution or shared runtime state.
    \item \textbf{Annotation-based API.} Developers can specify benchmark methods, parameters, setup, and teardown phases declaratively through annotations, making the benchmark implementations concise and less error-prone.
\end{itemize}

Despite the above-mentioned strengths, the execution model of JMH may have an important drawback.
Since each benchmark configuration is executed in isolation within its own JVM process, and since input parameters are fixed during that run, the benchmarked code is subject to what we call an execution in a \emph{sterile environment}.  
This means that a given method is typically invoked from a call site with identical parameters across all iterations, resulting in profiles and JIT optimizations that unlikely can be applied in a realistic application.
The opposite setting is an execution in a context-rich environment, i.e., methods are usually called passing different parameters, maybe even of different types.
As we show in our case studies, this execution model can lead to misleading conclusions, i.e., the performance behaviors observed in a sterile environment for a certain implementation may not match those observed in a context-rich environment.
Within realistic applications, code is commonly executed in a context-rich environment, rather than in a sterile one, in particular JCL classes and common libraries, as they are commonly used in multiple code locations of an application.

\section{Implications on Microbenchmarks that Lead to Misleading Results}\label{sec:implication}

A key observation in this work is that microbenchmarks execute code in what we call a \emph{sterile environment}, where the benchmarked code is typically invoked from a single call site, with a fixed parameter for each JVM run, and often as the only user of the functionality under evaluation.  
This execution style differs radically from real-world applications, where the code is used in multiple contexts, with heterogeneous parameters, and intermixed with other components.

The adaptive nature of the JIT compiler amplifies the problem.  
Optimizations such as inlining, loop unrolling, escape analysis, and scalar replacement are guided by runtime profiles, so the compiled machine code can vary depending on calling context, usage frequency, and input data.  
Microbenchmarks in sterile environments generate limited profiles, leading to machine code generated with a \emph{benchmark-tailored compilation}, which may differ from what would be generated in real workloads, which is more likely a \emph{general-use compilation}.  
This makes the performance observed in benchmarks unlikely to match that of realistic applications.

An additional implication of sterile benchmarking arises when comparing alternative implementations against classes from the JCL.  
Because the JVM executes significant portions of the JCL during initialization, many standard classes are already profiled and possibly JIT-compiled before the benchmark begins.  
By contrast, a custom implementation introduced only in the benchmark is evaluated in a sterile environment without such prior profiling.  
As we will show in our case studies, this asymmetry may result in profiles for the JCL class that are ``polluted'' by the code executed before the benchmark, while the alternative is profiled in isolation.
As we show in our case studies, this discrepancy can lead to wrong conclusions, misleading developers into preferring code that may actually perform worse in realistic scenarios.  

In summary, sterile environments are an inherent limitation of microbenchmarking: they produce simplified profiles that unlikely could be produced when the same code is executed within a real applications.  
While frameworks such as JMH mitigate common measurement errors, they cannot remove the fundamental divergence introduced by the execution within a sterile environment.  
This motivates the case studies in the next section, where we illustrate how these effects can lead to misleading performance results.  

\section{Experimentation Overview}\label{sec:overview}
Throughout the next sections, we analyze three case studies as common examples of microbenchmarks that can lead to misleading results.
In particular, we show that microbenchmarking a certain method has different performance behavior when executed in a sterile environment or in a more realistic setting.
Our realistic setting is obtained by exercising the benchmark methods with different parameters, therefore executing different code paths before the standard microbenchmark execution.

First (Section~\ref{sec:cs1}), we analyze the common scenario of optimizing a single small function, we use the hashcode computation as a use case.
Then (Section~\ref{sec:cs2}), we analyze the usage of library methods from the JCL. Here, we use the Stream API~\cite{jdk24streampackage} as a use case and we show that the same usage of the API, in a sterile environment and in a realistic setting lead to different performance.
Finally (Section~\ref{sec:cs3}), we analyze again the use of JCL classes, focusing on widely used collection classes from the \code{java.util} package, and we highlight a common mistake made by developers when implementing alternative implementations of those collections. We show that the performance of those collections may significantly differ when a benchmark uses a collection class from the JCL or its code clone.

\subsection{Evaluation Settings}
We perform the evaluation on an Intel machine with a 16-core Intel Xeon~Gold~6326 CPU~@2.9GHz with 256GB~RAM~@3.2GHz. The operating system is Ubuntu~22.04~LTS (kernel~5.15.0-25-generic).
Hyper-threading and turbo boost are disabled to ensure stable performance measurements.
We run our experiments on JDK-24 (Oracle JDK~24~\cite{oracle-jdk}, build~24.0.1+9-30).
For each case study, we make use the latest version of JMH at the time of writing (i.e., version~1.37), avoiding any well-known pitfall of JMH usage~\cite{jmh-do-dons-costa}. 
Moreover, each experiment is executed running 5 JVM runs, each of them with 5 warmup iterations of 5s (first 25s of execution are not measured) and 5 measured iterations of 5s, the reported results are the arithmetic mean of the measured iterations. 

\section{Case Study 1: Benchmarking a Single Function – hashCode()}\label{sec:cs1}

This case study focuses on a simple, but very common scenario: optimizing the performance of a short, stateless function.  
As an example, we show the computation of the hash code of a byte array. The  functionality is simple, easy to understand, widely used (e.g., to compute the hash codes of strings), and an actual subject of optimization attempts~\cite{loff2024vectorized,hashcode_dynatrace}.
Indeed, since Java~21, the JVM implements the methods for computing hash codes of primitive arrays as intrinsics using vector instructions.

In this case study, our baseline implementation corresponds to the hash function used by Java prior to this optimization.
We note that the goal is not to propose a new implementation of \code{hashCode()} that is faster than the aforementioned baseline. 
On the contrary, we aim to show a simple baseline next to a seemingly optimized alternative, which appears to provide better performance---but only when evaluated in a sterile environment. When evaluated in more realistic settings, it performs worse than the baseline.
To this end, we demonstrate how different execution strategies in micro-benchmarking can lead to misleading conclusions about which implementation is faster, and subsequently to wrong optimization decisions.

Figure~\ref{code:hashcode_baseline} shows the simple implementation used as the baseline. It contains a simple loop which computes the value of the \emph{polynomial rolling hash} function~\cite{10.5555/280635} shown in Equation~\ref{eq:prh}, where $s$ is an array with $N$ elements, $s[i]$ is the $i$-th element of $s$, $p$ and $m$ are positive integers. In the JCL, $p = 31$, $m = 2^{32}$, and the $\text{mod}~2^{32}$ operation is obtained implicitly due to overflows.

\begin{equation}
\label{eq:prh}
hash(s) = 
\begin{cases}
0 & \text{if $N=0$} \\
\sum^{N-1}_{i=0} s[i] \cdot p^{N-i-1}~~\text{mod}~m & \text{otherwise}
\end{cases}
\end{equation}

\begin{figure}
  \input{code/hashcode_baseline}
  \caption{Baseline implementation of hash code.}
  \label{code:hashcode_baseline}
  \Description[Baseline implementation of hash code]{Baseline implementation of hash code (polynomial rolling hashing function). A loop over the input byte array.}
\end{figure}

\begin{figure}
  \input{code/hashcode_variant}
  \caption{Alternative implementation of hash code.}
  \label{code:hashcode_variant}
  \Description[Alternative implementation of hash code]{Alternative implementation of hash code (polynomial rolling hashing function). A loop over the input byte array like the baseline only if the array length is greater than 32, otherwise the implementation uses a switch.}
\end{figure}

The alternative implementation in Figure~\ref{code:hashcode_variant} is (intended to be) optimized for short strings, which are considered important and optimized in other languages~\cite{small-strings-optimization}. For arrays with at most 32 elements, it computes the hash code using a switch, exploiting its fall-through semantic to minimize branches and to keep code size small.
We demonstrate the misleading results by evaluating the performance of the two implementations using three execution strategies.

\begin{figure*}
  \centering
  \includegraphics[width=\linewidth]{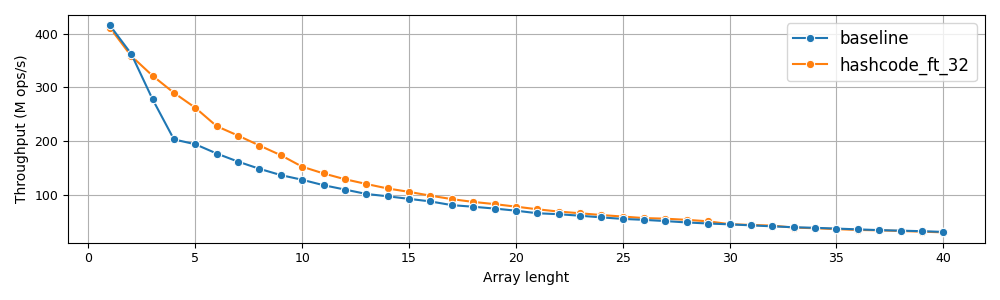}
  \caption{Throughput of \code{baseline} (Figure~\ref{code:hashcode_baseline})  and the variant, \code{hashcode\_ft\_32} (Figure~\ref{code:hashcode_variant}) using the \donothing execution strategy, \code{hashcode\_ft\_32} appears faster in a sterile environment.}
  \label{fig:eval:hashcode_naive}
  \Description{Throughput of baseline (Figure~\ref{code:hashcode_baseline}) and the variant, hashcode\_ft\_32 (Figure~\ref{code:hashcode_variant}) using the \donothing execution strategy, hashcode\_ft\_32 appears faster in a sterile environment.}

\end{figure*}

\paragraph{\donothing strategy.}  

The \donothing strategy corresponds to a widely used practice adopted by developers when they try optimize a function, i.e., using JMH to drive the optimization. With this strategy the developers do not take care of the narrow profiles that the JVM would collect when executing the benchmark repeatedly using always the same parameters, i.e., the benchmark is executed in the \emph{sterile environment} described in Section~\ref{sec:background:jmh}.  

Figure~\ref{fig:eval:hashcode_naive} shows the performance of the two \code{hashCode()} implementations evaluated using the \donothing execution strategy. 
The results are likely to lead to a conclusion that the alternative implementation is consistently faster than the baseline on arrays with length at most 32, and on par with the baseline afterwards (without any performance penalty).

However, this conclusion is misleading. When evaluated using the \donothing strategy, which in this particular case means that the \code{hashCode()} method is only called with arrays of single length in the scope of a single JMH run, the JVM is likely to generate machine code that is \emph{over specialized} to arrays of that particular length and does not reflect realistic usage of the method.

Before compiling a method to machine code, the JVM first collects profiling information including branch probabilities. In this case, the collected profile will indicate that only a single case of the switch is ever taken, while all other cases are never exercised.

Given such a (narrow) branch profile, the JVM can (and will) speculatively optimize the generated machine code to execute a code variant specialized for a specific array length if the input array length is the same. This is, however, always the case when evaluating performance using the \donothing execution strategy, leading to misleading results and wrong conclusion.

\begin{figure*}
  \centering
  \includegraphics[width=\linewidth]{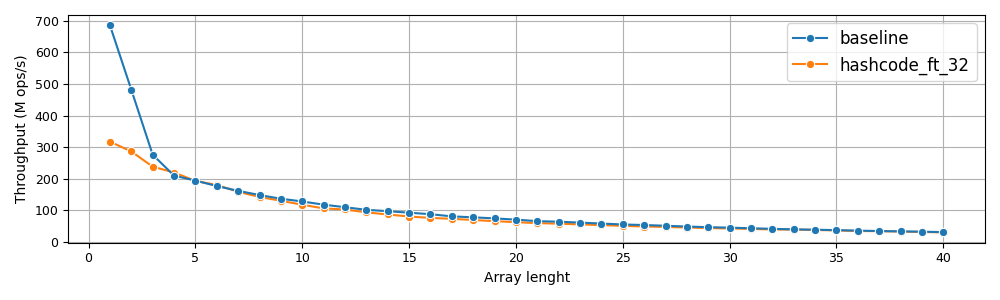}
  \caption{Throughput of \code{baseline} (Figure~\ref{code:hashcode_baseline}) and the variant, \code{hashcode\_ft\_32} (Figure~\ref{code:hashcode_variant}), using the \manualpollute execution strategy. After exercising all code paths during setup, \code{hashcode\_ft\_32} is slower.}
  \label{fig:eval:hashcode_improved}
  \Description{Throughput of \code{baseline} (Figure~\ref{code:hashcode_baseline}) and the variant, \code{hashcode\_ft\_32} (Figure~\ref{code:hashcode_variant}), using the \manualpollute execution strategy. After exercising all code paths during setup, \code{hashcode\_ft\_32} is slower.}

\end{figure*}

\paragraph{\manualpollute strategy.}  
To avoid the \emph{over specialization} caused by execution in a sterile environment, we propose an \manualpollute execution strategy leveraging JMH’s \texttt{@Setup} functionality, which is a user-defined method executed once before the repeated execution of the benchmark.
Before measuring performance, we exercise the benchmark method with a set of input parameters designed to cover all code paths of the \code{hashCode()} implementation.
This ensures that the JIT compiler observes less narrow profiles and applies optimizations that better reflect realistic usages.

Figure~\ref{fig:eval:hashcode_improved} shows the performance of the two \code{hashCode()} implementations with the \manualpollute execution strategy. 
Measured using the \manualpollute execution strategy, the alternative implementation is no longer faster; on the contrary, it performs worse than the baseline on most of the data points.
This is because the alternative implementation no longer benefits from the over specialization of the switch statement happening under the \donothing execution strategy---exercising diverse code paths results in a profile that causes the JVM JIT compiler to produce machine code that always needs to be able to deal with different input array lengths instead of a single one.
This highlights the need for exercising diverse paths in order to obtain representative benchmark results.

Table~\ref{tab:hashcode:stats} shows the statistics of the speedup factors of the alternative implementation against the baseline implementation for the execution strategies \donothing and \manualpollute. As reported in the table, with the \donothing strategy the alternative implementation shows speedup factors ranging from 0.97x to 1.43x, with a geometric mean of 1.10 (i.e., it seems to be a candidate optimization), while with the \manualpollute strategy it shows speedup factors ranging from 0.46x to 1.05x, with a geometric mean of 0.91 (i.e., it is slower on most of the array lenghts).

\begin{table}
     \caption{Statistics of the speedup factors of \code{hashcode\_ft\_32} (Figure~\ref{code:hashcode_variant}) against \code{baseline} (Figure~\ref{code:hashcode_baseline})  using the strategies \donothing and \manualpollute.}
   \centering
    \begin{tabular}{lllrrr}
    \toprule
    Strategy & Min & Max & GeoMean \\
    \midrule
    \donothing & 0.97x & 1.43x & 1.10x \\
    \manualpollute & 0.46x & 1.05x & 0.91x \\
    \bottomrule
    \end{tabular}
    \label{tab:hashcode:stats}
\end{table}

\begin{figure}
  \centering
  \includegraphics[width=\linewidth]{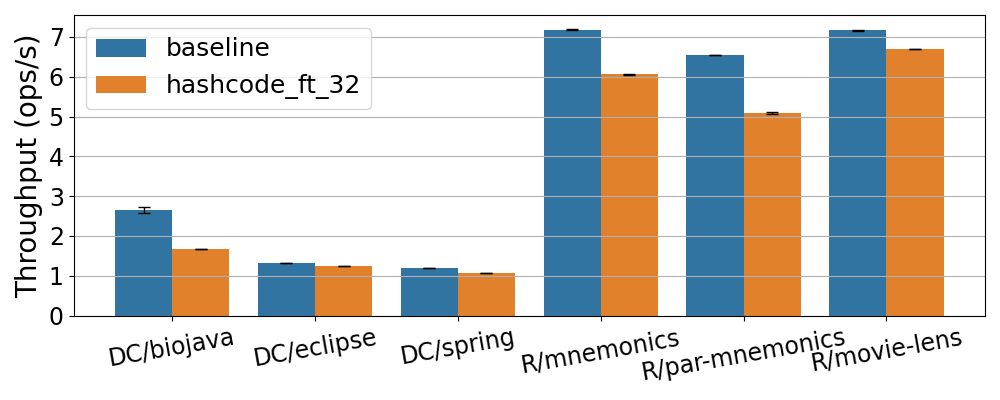}
  \caption{Realistic traces on Dacapo-Chopin (DC) and Renaissance (R). The benchmark confirms the results of the \manualpollute strategy. Error bars report the confidence interval.}
  \label{fig:eval:hashcode_traces}
  \Description{Realistic traces on Dacapo-Chopin (DC) and Renaissance (R). The benchmark confirms the results of the \manualpollute strategy.}
\end{figure}

\paragraph{Realistic Traces.}  
Finally, we validate that the \manualpollute strategy provides representative results using a record-and-replay approach~\cite{chandy2009rollback,chen2015deterministic}.
We profile the execution of two widely used Java benchmarks, DaCapo-Chopin~\cite{dacapo-chopin} and Renaissance~\cite{renaissance}, collecting traces of \code{hashCode()} invocations and recording the lengths of the input arrays.\footnote{We note that trace replay is executed in a single thread and that the trace always represents a total order of the \code{hashCode()} invocations.}
Then, we selected the three traces with a higher number of invocations to \code{hashCode} for each benchmarks and we replay them in a controlled JMH benchmark. 
We used such a record-and-replay approach because within these large benchmark suites only a rather small fraction of the total execution time is spent within the \code{hashCode()} method, making it difficult to measure the relative performance difference of the two implementations.

Figure~\ref{fig:eval:hashcode_traces} shows the performance of the two \code{hashCode()} implementations with the realistic-traces execution strategy.
The performance results obtained with realistic traces are in line with those of the \manualpollute strategy, confirming that pre-exercising the function with diverse inputs to cover all the execution paths yields results consistent with actual workloads.
While the realistic-traces approach is likely the most precise one to obtain representative insights, it is significantly more complex and time-consuming to implement (i.e., it requires the implementation of a suitable record-and-replay infrastructure) compared to the \manualpollute strategy.

\paragraph{Takeaway}
This case study demonstrates that microbenchmarks executed in sterile environments can easily lead to misleading results, even in the case of a seemingly simple task of optimizing a small stateless function.
The \donothing approach---repeatedly invoking a function with fixed inputs---is a common practice in code optimization but often leads to wrong conclusions.  
On the other hand, exercising the function across its execution paths during benchmark setup produces results that are consistent with realistic traces, thereby avoiding incorrect optimization choices.  
We recommend that developers always adopt this practice.
\section{Case Study 2:  Benchmarking Libraries – Stream API}\label{sec:cs2}

The second case study focuses on a more complex scenario involving performance evaluation of application code that uses the Stream API to process data in collection classes.

The Stream API was introduced in Java~8 and provides a monadic-like interface to Java collections that allows processing data in a functional and declarative way.
Streams abstract away explicit iteration and represent computations as pipelines of operations that transform  individual elements, relying on lambda expressions and method references for operation composition.

While the abstractions offered by the Stream API provide clear benefits to the developer, the compiler is faced with fragmented code that relies on numerous invocations of virtual (interface) methods representing user callbacks.
Efficiently implementing these invocations is key to high performance, and the JIT compiler can attempt to optimize them, e.g., using polymorphic inline caches, or by speculative inlining specific callees.

Performance of stream-based code therefore heavily depends on profile-guided optimizations, which makes the Stream API a great candidate for studying the impact of execution environment on the performance observed during micro-benchmarking experiments.

As a use case, we employ the JEDI~\cite{jedi} benchmark, that uses S2S~\cite{s2s} to translate queries from the TPC-H benchmark~\cite{tpch}, a widely-used benchmark for relational databases.

In contrast to the previous case study, we do not compare different query implementations; instead, we show that even when executing identical implementations, performance results can change significantly depending on whether the Stream API has been exercised before the benchmark execution or not.

\paragraph{\donothing strategy.}  
The \donothing strategy corresponds to the standard JMH execution: each benchmark (a single TPC-H query) is executed in isolation, starting from a fresh JVM process, with no prior use of the Stream API.
This execution in a sterile environment leads the JVM to collect execution profiles based solely on the specific query.
As a result, speculative optimizations become highly specialized, as each call site always observes the same lambda function.
This specialization often leads to optimistic inlining decisions and optimized execution paths that do not generalize to realistic workloads in which the Stream API is going to be called from many different places with different callbacks.

\paragraph{\manualpollute strategy.}  
To avoid such over specialization, we propose a \manualpollute strategy similar to the one used in the previous case study, by exploiting JMH’s \texttt{@Setup} feature.
In this case, however, we do not aim to exhaustively exercise all possible code paths of the Stream API (which would be more difficult given the size of the library).
Instead, we pre-execute a representative set~\cite{on-the-wild-khatchadourian,on-the-wild-costa,on-the-wild-eduardo} of Stream operations (e.g., \code{map}, \code{filter}, and \code{collect}) during the setup phase.
This simple strategy is sufficient to alter the profiles collected by the JVM, leading to less specialized machine code for the subsequent execution of the benchmarked query.

\begin{figure*}
	\centering
	\includegraphics[width=\linewidth]{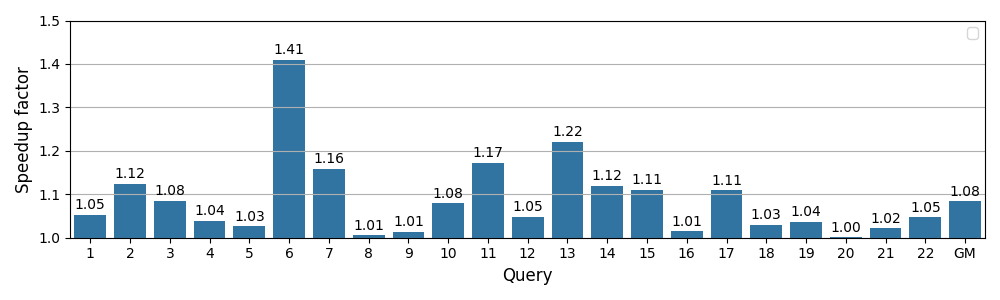}
	\caption{Speedup of the \donothing execution strategy over the \manualpollute one for the JEDI~\cite{jedi} benchmarks. Values greater than $1.0$ indicate that in a sterile environment the benchmark executes faster. The last bar (GM) reports the geometric mean of the speedups on all queries.}
	\label{fig:eval:stream_speedup}
    \Description{Speedup of the \donothing execution strategy over the \manualpollute one for the JEDI benchmarks. Values greater than $1.0$ indicate that in a sterile environment the benchmark executes faster. The last bar (GM) reports the geometric mean of the speedups on all queries.}
\end{figure*}

\paragraph{Results.}  
Figure~\ref{fig:eval:stream_speedup} compares the performance
of different stream-based queries observed under the \donothing and \manualpollute execution strategies, showing the speedup factor of the \donothing strategy with respect to the \manualpollute strategy.
The queries executed under the \donothing strategy are always faster, with an average speedup of 1.08$\times$ (geomean) and reaching the maximum of 1.41$\times$ (Query~6).
This clearly demonstrates that the sterile environment leads to an unrealistic performance advantage, as the machine code produced in this setting is specialized for the limited profiles observed during compilation.
By contrast, even a modest pre-execution of Stream operations in the setup phase is enough to invalidate these unrealistic specializations, leading to slower but more representative performance measurements.  

Because the scenario covered in this case study is more complex than the previous one, as it involves a large library, we have manually analyzed the compilation logs of several queries, comparing those obtained with the \donothing strategy and the \manualpollute strategy.
Ideally, all the lambdas should be inlined into the stream API methods (i.e., the callees), but only in the scope of the callers (i.e., the query implementations in this case study) so that the code specialization stays local to the caller.

Indeed, we observed that with the \donothing strategy, many functions passed as lambdas to the Stream API within the query implementations, e.g., the predicates passed to the \code{filter()} stream method, were speculatively inlined within the compiled code of the Stream API itself. 
With the \manualpollute strategy, and likely with any application that makes use of the Stream API in multiple code locations, such a speculative optimization would normally not be applied, since the profiles of the Stream API methods would show multiple different lambdas passed as parameters.
Moreover, with the \manualpollute strategy, the JVM even tried to inline some of the lambdas into the stream API methods, but since the Stream API methods are not inlined into the callers some specializations became global and applied to all callers in an inefficient way.

\paragraph{Takeaway.}  
This case study demonstrates that the issue of performance evaluation in a sterile environment also affects also complex libraries such as the Stream API.
We have shown that executing identical code in a sterile environment and in an environment obtained after modestly exercising the library can lead to significant performance differences, in this case up to 1.41$\times$.
Developers and researchers should be cautious when benchmarking library methods in isolation, as the conclusions drawn from such benchmarks may not reflect realistic usage of the library.

\section{Case Study 3:  Benchmarking Libraries Used during JVM Initialization – Collections}\label{sec:cs3}

The third case study focuses on a subtle but common mistake when benchmarking alternative implementations to JCL classes.
Some of these classes, such as Java collections offered in the \code{java.util} package (e.g., \code{ArrayList} and \code{HashMap}) are not only essential for application developers, but are also heavily exercised by the JVM itself during the initialization phase, which takes place before the execution of any application/benchmark code.
As a consequence, their methods are profiled and possibly already compiled by the JIT compiler before application/benchmark code begins execution.

This behavior has an important implication for benchmarking.
We focus our case study on Java collections, since they are widely adopted in most Java applications.
When developers evaluate alternative collection implementations---an effort found in both research papers~\cite{steindorfer2015optimizing,prokopec2017making,nerella2014exploring} and popular third-party libraries such as Eclipse Collections~\cite{eclipse-collections}---the comparison of an alternative collection with a JCL one is typically performed using microbenchmarks~\cite{collectionbench}.  
However, while the user-defined collections are executed in a sterile environment, the JCL classes have already been exercised during JVM initialization, even when using the \donothing strategy, since the JVM initialization always takes place. 
This introduces a systematic bias: JCL classes may appear slower than they are, because they have been profiled during the execution of initialization code which is unrelated to the benchmark, while the alternatives benefit from the execution in a sterile environment and get optimized specifically for the microbenchmark.

We demonstrate this effect in an experiment in which we compare the performance of selected JCL collection classes and copies of the classes (i.e., code clones) placed in a different package.
The only difference between the two sets of classes is that the JCL versions are exercised during JVM initialization, while the copies are not.
We evaluate these implementations using CollectionBench~\cite{collectionbench}, an existing benchmark suite for Java collections implemented as a set of JMH microbenchmarks.
In particular, we focus on the benchmarks designed for the HashMap and the HashSet JCL classes. 
Both benchmarks for each class are composed of 4 workloads: 
\begin{itemize}
    \item \code{CONTAINS} (\code{contains()} method).
    \item \code{COPY} (\code{HashMap.putAll() and \code{HashSet.addAll()}} methods).
    \item \code{ITERATE} (iteration using iterators methods).
    \item \code{POPULATE} (\code{HashMap.put()} and \code{HashSet.add()} methods).
\end{itemize}

\begin{table}
\caption{Statistics of the speedup factors of the JCL code- clones against the original ones for the JDKMaps (left) and JDKSets (right) benchmarks in CollectionBench.}\vspace{-3mm}
\begin{tabular}{l|lll|lll}
\hline
& \multicolumn{3}{c|}{JDKMap} & \multicolumn{3}{c}{JDKSet} \\
Workload & Min & Max & GM & Min & Max & GM \\
\hline
CONTAINS & 1.0x & 1.21x & 1.05x & 1.0x & 1.22x & 1.06x \\ 
COPY & 1.12x & 1.55x & 1.34x & 1.06x & 1.25x & 1.16x \\ 
ITERATE & 0.99x & 1.1x & 1.01x & 0.98x & 1.01x & 0.99x \\ 
POPULATE & 1.08x & 1.47x & 1.19x & 1.09x & 1.54x & 1.26x \\ 
\bottomrule
\end{tabular}
\label{tab:stats:collections}
\end{table}

\paragraph{Results.}  

Table~\ref{tab:stats:collections} summarizes the statistics of the speedup factors for the \code{HashMap} (left) and the \code{HashSet} (right) classes.
Notably, the copies consistently outperform the original JCL classes when executed using the \donothing strategy.  
The copied classes show speedup factors ranging from 0.99x (ITERATE benchmark) to 1.55x (COPY benchmark) on \code{HashMap} and from 0.98x (ITERATE benchmark) to 1.54x (POPULATE benchmark) on  \code{HashSet}.

We note that for this experiment we do not need to use the \manualpollute strategy, and we use the \donothing.
Indeed, we use the JVM initialization as a realistic (and unavoidable) phase that exercises code paths in the collection classes, so the collected profiles are inherently polluted, whereas the copies of the JCL classes are executed in a sterile environment.
Therefore, the performance difference originates entirely from the JVM initialization

\paragraph{Takeaway.}  
This case study highlights the unfairness of comparing JCL collections (or, more generally, any class used during the JVM initialization) with user-defined alternatives that are not used during the JVM initialization. 
The systematic bias introduced by JVM initialization can easily mislead developers and researchers to underestimate the performance of JCL classes and overestimate the benefits of their alternatives. 
As a solution, we suggest two practical ways to ensure a fair performance comparison:
\begin{enumerate}
	\item \textbf{Copying JCL classes.} This ensures that both the JCL version and the alternatives are executed in the same (sterile) environment, leading to a fair comparison for a given workload. However, such approach is not ideal setting, because even though it removes the aforementioned bias, it still suffers from the problem of micro-benchmarking in a sterile environment. This can be mitigated by introducing code exercising the classes in the @Setup phase, as discussed in the prior case studies, but such usage may be different from the real usage during JVM initialization.
	\item \textbf{Using Java patch mechanisms.}~\cite{loff2024vectorized} A more robust approach is to replace JCL classes with user-defined implementations, ensuring that both versions are exercised during initialization and benchmarked under realistic conditions. However, such an approach can be applied only if the alternative implementations are fully compatible with the replaced JCL classes, i.e., including non-public interfaces and contracts relied upon by other JCL classes.
\end{enumerate}

\section{Related Work}\label{sec:relwork}
Our work builds upon a well-established body of research highlighting the perils of performance measurement and assumes a baseline methodology grounded in current best practices. 
These cover aspects such as experimental and statistical rigor~\cite{georges2007statistically,blackburn2008wake,kalibera2013rigorous,blackburn2016truth}, balancing accuracy against execution cost~\cite{kalibera2013rigorous,japke2025muoptime}, and accounting for external factors such as memory layout or noisy neighbors that can distort results~\cite{mytkowicz2009producing,laaber2019software}.
The assumed baseline methodology also accounts for basic effects of warmup on performance, especially on the JVM and other managed-language runtimes featuring a garbage collector and a JIT compiler. 
Even though benchmark performance can oscillate indefinitely~\cite{barrett2017virtual}, the practical goal is to prevent measurements from the early phases of benchmark execution (when most compilation and other performance-disrupting changes take place) from polluting the evaluation data. 
The boundary between warmup and the (practically relevant) steady state is usually established through experience or visual inspection of performance data~\cite{traini2022effective}, with recent work exploring AI-based methods for automating the detection of the end of the warmup phase~\cite{traini2024ai-driven,trovato2025amber}.

Beyond statistical rigor and warmup effects, challenges lie in the design of microbenchmarks, which requires deep understanding of hardware and software platform details, e.g., to counteract optimizations enabled by the dynamic nature of the JVM~\cite{goetz2004dynamic,shipilev2013jmh}. 
Early case studies showed that even simple benchmarks can lead to misleading conclusions and identified numerous methodological and design pitfalls~\cite{gil2011microbenchmark,shipilev2014nanotrusting,ponge2014avoiding,horky2015conducting}.
Frameworks such as JMH~\cite{jmh} provide the necessary tools to address many of these pitfalls, yet developers still struggle to use them effectively in practice~\cite{jmh-do-dons-costa}.

Our work extends this line of research by highlighting the impact of benchmark execution context on observed performance. 
Specifically, we show that ignoring the effects of benchmark execution on the profiles used by the JIT compiler can result in aggressive optimizations that would not occur in a real application, leading to unrealistic performance. 
We argue that controlling for the execution environment must encompass not only low-level factors like memory layout, but also high-level platform behaviors that emerge only in realistic contexts. 
In this sense, our study extends earlier lessons on flawed benchmark design~\cite{gil2011microbenchmark}, by demonstrating that even carefully constructed microbenchmarks may still misrepresent application performance if the execution context is ignored.

\section{Conclusion}\label{sec:end}

Microbenchmarking is a powerful technique to analyze the performance of code snippets and libraries, but we showed that it is also prone to pitfalls. 
Benchmarks executed in sterile environments, where a function is invoked repeatedly with fixed inputs, can easily lead to misleading results due to unrealistic specialization performed by the JIT compiler. 
This can result in wrong conclusions about the effectiveness of optimizations or the relative performance of different implementations.
With three case studies we demonstrated this issue. 
First, we showed how a variant of the \code{hashCode()} function appeared faster in sterile settings but slower under realistic conditions.
Second, we illustrated how the performance of the Stream API can vary depending on whether the library has been exercised before benchmarking. Finally, we analyzed the widely used collections \code{HashMap} and \code{HashSet}, showing that JCL classes suffers from profiling information collected during JVM initialization, leading to unfair comparisons against alternative implementations.

As future work, we foresee the development of a record-and-replay framework that automates the generation of realistic microbenchmarks by capturing selected usages of a feature of interest. A major challenge for this work is capturing only the minimal information required to faithfully reproduce the application.

\begin{acks}
This work has been supported by the Hasler Fundation (2025-03-13-355) and the Swiss National Science Foundation (IZSEZ0\_229176).
\end{acks}

\printbibliography

\end{document}